# A POSSIBLE EXPLANATION OF THE VOID IN THE PYRAMID OF KHUFU ON THE BASIS OF THE PYRAMID TEXTS

Giulio Magli

Politecnico di Milano, Milan, Italy.

giulio.magli@polimi.it

A recent exploration has shown the presence of a significant void in the pyramid of Khufu at Giza. A possible explanation of this space, interpreted as a chamber connected to the lower north channel and aimed to contain a specific funerary equipment is tentatively proposed. According to the Pyramid Texts, this equipment might consist of a "Iron throne", actually a wooden throne endowed with meteoritic Iron sheets.

KEYWORDS: Pyramid Texts – Khufu void – Archaeoastronomy of ancient Egypt.

## 1. INTRODUCTION

This paper concerns the internal structure of the Pyramid of Khufu at Giza. In what follows it is assumed that the reader has some familiarity with this monument (for a discussion see e.g. Lehner 1998). The origin of the present work is the recent investigation carried out with muon detectors, which has shown that the Pyramid of Khufu does not – in spite of repeated claims -contain inhomogeneities and/or vacuum zones of a certain relevance in the bulk of the stone material, with only one exception (Morishima et al. 2017). Homogeneity is a result which was to be expected, taking into account the extreme care exhibited by its builders in any detail of construction. The unique exception to the structural homogeneity of the monument which has been discovered is a huge void located exactly along the midway north-south section of the pyramid, over the Great Gallery. The void has a cross section similar to the Grand Gallery and a length of 30 m minimum. The muon excess is similar to the one generated by the Grand Gallery itself, which means that the volume of the two voids is of the same order. In other words, the newly discovered space is certainly a chamber or a system of chambers. The center of the void is located between 40 m and 50 m from the floor of the Queen's Chamber.

Although the detailed structure has yet to be determined, it is obvious that this void cannot be interpreted as a result of sloppy behavior on the part of the builders. Indeed, the very fact that this is the *unique* relevant non-homogeneity present in the bulk of the pyramid points to its deliberateness. Further, the void cannot be interpreted – as proposed by some Egyptologists - as a chamber devoted to relief weight from the roof of the Great Gallery. In fact, in all structural situations in which a relief of the weight exerted on roofs was needed, it was dutifully operated by the ancient Egyptians. Specifically, the Khufu architect adopted the inverted V vault for the roof of the Queen Chamber and for the roof of the uppermost chamber located over the King's chamber (for a detailed discussion see Magli 2013). The roof of the Great Gallery was of course in danger as well, and for this reason the builders adopted the corbelled ceiling for it. The uppermost part of the corbel is closed by flat slabs, but – even admitting the existence of a relieving tunnel for this last part – this structure would be a small corridor running immediately over the ceiling, like the one discovered some years ago in Meidum, and certainly not a huge space such as the one shown by the prospection.

All in all, if the void cannot be attributed either to chance or to structural reasons, its presence must be functional to the scope of the pyramid: the tomb of the Pharaoh, built to assure its eternal afterlife. Thus, there must be a solid reason for its existence, and this reason must be deeply connected with the Egyptian funerary religion. The present paper is a first, admittedly highly speculative, attempt at possible explanation on the basis of the funerary beliefs of the epoch.

## 2. THE PYRAMID TEXTS "OF RESURRECTION" AND THE INTERIOR CHAMBERS IN KHUFU'S PYRAMID

The Great Pyramid is endowed with architectural elements which are purely symbolic, related only to the ideas on the afterlife. In particular, there are four long and complex shafts which cannot be accessed by human beings for the simple reason that they are two small. They start from the Queen and from the King's chamber and point towards the north and south faces. The upper channels, coming from the King's chamber, exit on the faces, the lower channels stop on suitably built doors endowed with copper handles. Four out of four channels are aimed to the stars mentioned in the funerary spells known as Pyramid Texts (PT) as destinations in the sky for the dead king: the circumpolar stars in the north, Sirius and Orion at culmination in the south (see Magli and Belmonte 2009 for an accurate discussion). The PT, discovered by Gaston Maspero at the end of the 19th century, are written in the funerary chambers of pyramids constructed around 200 years after those of Giza, but Egyptologists agree that they must be much more ancient than that; they therefore can furnish us a formidable key to the interpretation of the symbolic elements of the pyramids. Besides the orientation of the channels, another example is the very presence of the doors at the end of the Queen's chamber shafts. Indeed, they probably are a materialization of those "doors of the sky" which are frequently cited in the texts, and it is extremely likely that the Queen's chamber in itself is the place where a statue of Khufu was placed, in order to perform on it the ceremony of the Opening of the Mouth, described in the PT and essential for making the "spirit" of the Pharaoh able to survive death and to go in the sky crossing these doors. Finally, only with the obsessive interest for the circumpolar stars – continuously cited in the PT as the "imperishable" stars - can the maniacal precision by which the pyramid was oriented be explained.

In the present paper, I propose that the PT may help us in understanding what is located in the void, and therefore, why it was constructed. Let us indeed read one (PT 536) of the many utterances usually called by Egyptology "resurrection texts" (Faulkner 1998):

> *"The doors of the sky are opened for you, the doors of Nut are thrown opened for*
> *you, the doors of the firmament are thrown opened for you. Endure, says Isis, in*
> *peace, says Nepthis, when they see their brother. Raise yourself, loose your*
> *bonds, throw off your dust, sit on this your iron throne."*

It appears therefore that, after crossing the "doors of the sky" the king will seat on an "iron throne". Many other, similar utterances mention the same process, and some also specify that the king will sit on the throne after crossing specifically the *northern* doors and that a *stairway* gives access to the throne. For instance, again in PT536 we read "may the honoured ones clap their hands at the stairway of your throne".

Observe now that it is not uncommon for objects mentioned in the PT to be documented in the archaeological records. An important example are the instruments mentioned in the opening of the mouth ceremony (a forked knife, an adze, a serpent-like knife) which are all known archaeologically. May the iron throne have been a real object in Khufu's burial equipment?

Interestingly, we do know how a throne of Khufu times looked like, because the throne of Khufu's mother, Queen Hetepheres, was recovered in fragments in the tomb containing her funerary equipment found near her son's pyramid (Der Manuelian 2017). It is a low chair made of cedar with faience inlays, gilded and endowed with teen gold sheets. Thus, Khufu Iron throne might have been a similar object: a cedar chair embellished with inserts of Iron sheets instead of gold. Of course, one should not think to melted Iron, which was not in use at those times, but only to the rare Iron coming from the sky in the form of Iron meteorites. Small objects made of this kind of Iron (recognizable from the high Nickel percentage) are known since pre-dynastic times and were used for the Pharaoh's burial equipments, as recently confirmed by the analysis of Tutankhamen sword (Comelli at al 2016) (in this connection it may be noted that a thin plate of Iron was actually found in the explorations of the 19[th] century of the north face of the Pyramid of Khufu, but it turned out to be not of meteoritic origin and therefore cannot belong to Khufu times).

If Khufu's Iron throne is really located in the newly discovered chamber, it was put in place during the construction of the pyramid and never accessed since. It may well be located at the central end of the void, which, according to the prospection, lies directly on the vertical axis from the apex of the pyramid, crossing the end of the Great Gallery, the so-called "big step" and the Queen's chamber underneath. The chamber might thus be a non-functional copy of the Great Gallery, with an ascending staircase and the throne at the uppermost end. However, if we admit this hypothesis, how was Khufu spirit meant to access the chamber?

The answer seems quite natural: trough a symbolic door, precisely the door located at the end of the northern lower shaft. This shaft has been explored two times; the first time, by Rudolph Gantenbrinck, whose robot however was unable to negotiate a bend which occurs after a few meters, and a second time in a an exploration carried out by National Geographic. The results of this second exploration have never been published, but news reports show that the team was able to reach the end of the shaft and photographed a door analogue to the one of the southern shaft; they also said that the length of the shaft is roughly equal to that of the other lower shaft. Basing on this scant information, the end of the shaft can be estimated to be over the beginning of the Great Gallery, as tentatively shown in Fig. 1. It may, therefore, be connected with the lowermost end of the sealed chamber.

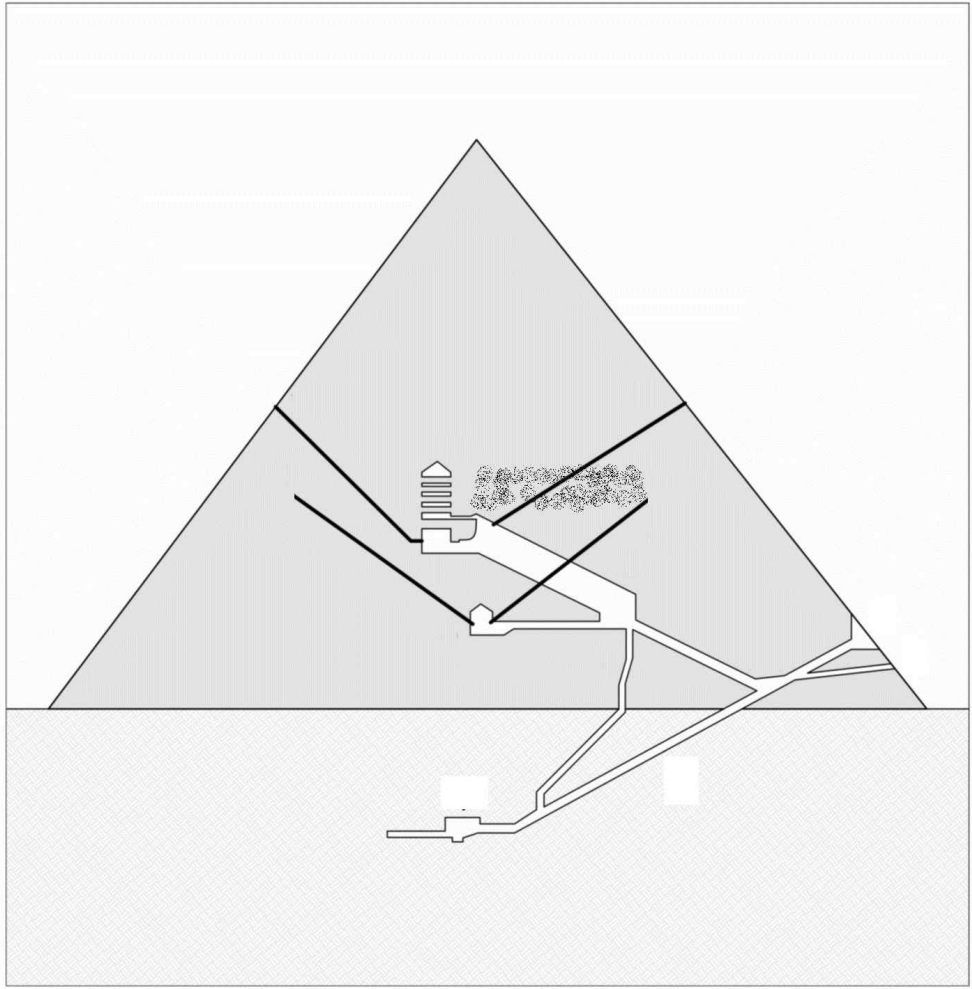

**Figure 1.** North-south section of the Great Pyramid showing as a dust-filled area the hypothetical extent of the "void" chamber, in connection with the lower southern shaft. The upper southern shaft does not intersects the hypothetical chamber (as instead suggested by the 2-dimensional section) because it is displaced to the west with respect to the Great Gallery.

### 3. CONCLUSIONS

For the moment, the prospections are too approximate to allow us any definitive conclusion; however, the existing information – together with what we know about the funerary religion of ancient Egypt – are sufficient to attempt at an explanation of the void which has been shown to exist inside the Pyramid of Khufu. It appears indeed that this void is not a failure in the construction, neither can be interpreted as a structural feature such as a reliving chamber. We proposed here that the void corresponds to a non functional "copy" of the Great Gallery beginning at the egress of the northern lower shaft and built to contain a symbolic object located under the apex of the pyramid. This object might be a throne endowed with sheets of meteoritic Iron, in accordance with some "resurrection" passages of the Pyramid Texts.

The author of the present paper is well aware that this theory is highly speculative. However, it has a bonus which pseudo-scientific theories usually have not. The possibility of being falsified by a – long sought, already since much before the discovery of the void – new exploration of the northern lower shaft.